# Transport Properties of the Lorentz Gas in Terms of Periodic Orbits

**PREDRAG CVITANOVIĆ**

Niels Bohr Institute, Blegdamsvej 17, DK-2100 Copenhagen Ø, Denmark

**JEAN-PIERRE ECKMANN**

Département de Physique Théorique, Université de Genève, CH-1211 Geneva 4, Switzerland

and

**PIERRE GASPARD**

Service de Chimie-Physique, Université Libre de Bruxelles, Campus Plaine, B-1050 Brussels, Belgium

**Abstract**—We establish a formula relating global diffusion in a space periodic dynamical system to cycles in the elementary cell which tiles the space under translations.

## Introduction

The diffusive properties of a "Lorentz gas" [L] have been studied extensively in the literature, see refs. [S, BS, Bu, GN, MZ]. The novelty of the approach presented here is that it provides an explicit connection between the global diffusion and the dynamics restricted to an elementary cell. Our method applies to any hyperbolic dynamical system that is a periodic tiling $\widehat{M} = \bigcup_{\hat{n} \in T} M_{\hat{n}}$ of the dynamical phase space $\widehat{M}$ by *translates* $M_{\hat{n}}$ of an *elementary cell* $M$, with $T$ the Abelian group of lattice translations. Furthermore, each elementary cell may be built from a *fundamental domain* $\widetilde{M}$ by the action of a discrete (not necessarily Abelian) group $G$.

These concepts are best illustrated by a specific example, a Lorentz gas based on a Sinai billiard [S] on the hexagonal lattice as in Fig. 1, with disks sufficiently large so that no infinite length free flight is possible.

It should be stressed that $\widehat{M}$ refers here to the full phase space, i.e., both the spatial coordinates and the momenta. The spatial component of $\widehat{M}$ is the complement of the disks in the *whole* space. We shall relate the dynamics in $M$ to diffusive properties of the Lorentz gas in $\widehat{M}$, using functional determinants and $\zeta$-functions.

It is convenient to define a time evolution operator for each of the 3 cases of Fig. 1. $\hat{x}_t = \hat{\phi}^t(\hat{x})$ denotes the point in the global space $\widehat{M}$ obtained by the flow in time $t$. $x_t = \phi^t(x)$ denotes the corresponding flow in the elementary cell; the two are related by

$$\hat{n}_t(x) = \hat{\phi}^t(x) - \phi^t(x) \in T \,, \tag{1.1}$$

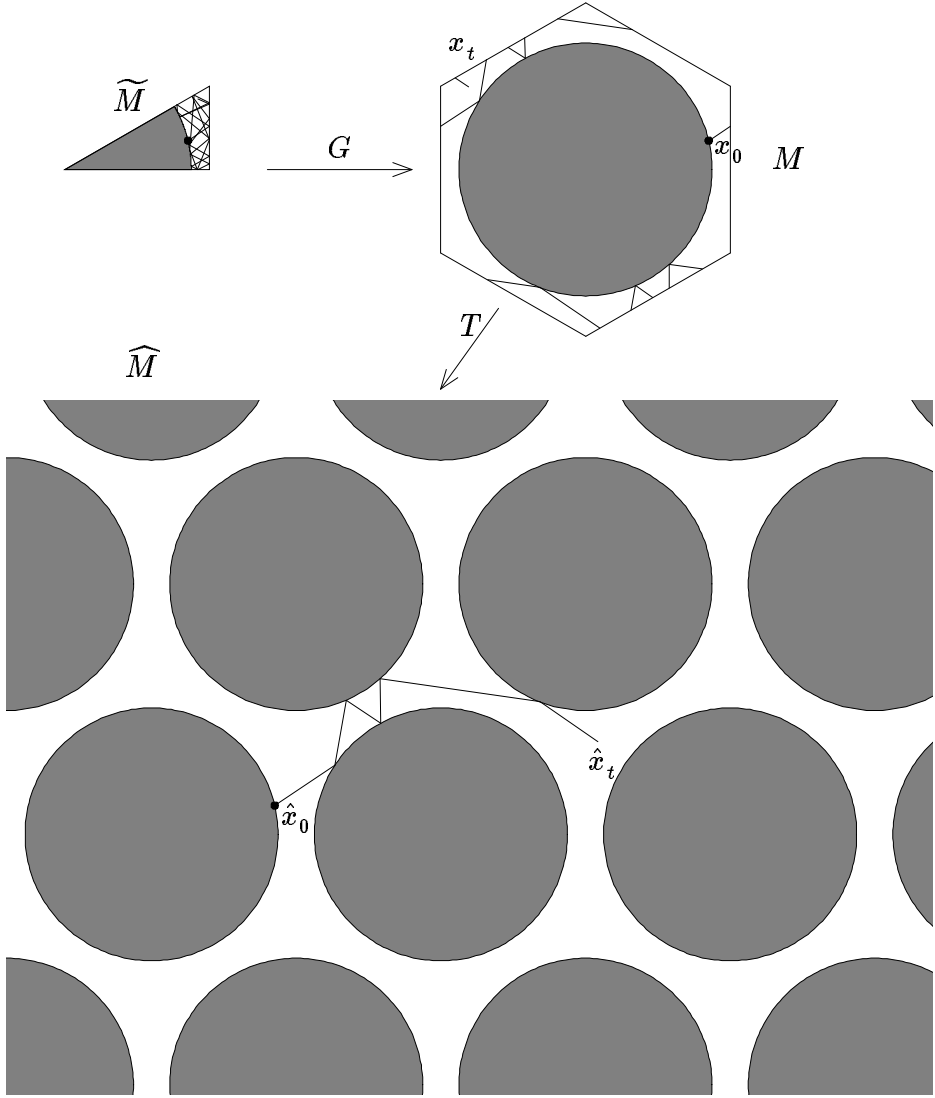

**Fig. 1**. Tiling of $\widehat{M}$, a periodic lattice of reflecting disks, by the fundamental domain $\widetilde{M}$. Indicated is an example of a global trajectory $\hat{x}_t$ together with the corresponding elementary cell trajectory $x_t$ and the fundamental domain trajectory $\tilde{x}_t$.

the translation of the endpoint of the global path into the elementary cell $M$. The quantity $\tilde{x}_t = \tilde{\phi}^t(\tilde{x})$ denotes the flow in the fundamental domain $\widetilde{M}$; $\tilde{\phi}^t(\tilde{x})$ is related to $\phi^t(\tilde{x})$ by a discrete symmetry $g \in G$ which maps $\tilde{x}_t \in \widetilde{M}$ to $x_t \in M$.

Fix a vector $\beta \in \mathbf{R}^d$, where $d$ is the dimension of the phase space. We will compute the diffusive properties of the Lorentz gas from the generating function

$$\langle e^{\beta \cdot (\hat{x}_t - x)} \rangle_M \,, \tag{1.2}$$

where the average is over all $x \in M$.

The diffusive properties follow by studying

$$Q(\beta) = \lim_{t \to \infty} \frac{1}{t} \log \langle e^{\beta \cdot (\hat{x}_t - x)} \rangle_M \,, \tag{1.3}$$

and its derivatives at $\beta = 0$. Clearly $Q(0) = 0$, and if by symmetry all odd derivatives vanish, there is no drift

$$\left. \frac{\partial}{\partial \beta_i} Q(\beta) \right|_{\beta=0} = \lim_{t \to \infty} \frac{1}{t} \langle (\hat{x}_t - x)_i \rangle_M = 0 \,. \tag{1.4}$$

In that case, the second derivatives

$$\frac{\partial}{\partial \beta_i} \frac{\partial}{\partial \beta_j} Q(\beta)\bigg|_{\beta=0} = \lim_{t\to\infty} \frac{1}{t} \langle (\hat{x}_t - x)_i (\hat{x}_t - x)_j \rangle_M , \tag{1.5}$$

yield a (generally anisotropic) diffusion matrix. The spatial diffusion constant is then given by

$$D = \frac{1}{2} \sum_i \frac{\partial^2}{\partial \beta_i^2} Q(\beta)\bigg|_{\beta=0} = \lim_{t\to\infty} \frac{1}{2t} \langle (\hat{q}_t - q)^2 \rangle_M , \tag{1.6}$$

where the $i$ sum is restricted to the spatial components $q_i$ of the phase space vectors $x$.

We next describe the connection between Eq.(1.3) and periodic orbits in the fundamental cell. As the full $\widehat{M} \to \widetilde{M}$ reduction is complicated by the nonabelian nature of $G$, we first introduce the main ideas in the abelian $\widehat{M} \to M$ context, and discuss the problems associated with the full reduction in Sect. 3.

## Reduction from $\widehat{M}$ to $M$

The general idea of our approach consists in writing

$$\langle e^{\beta \cdot (\hat{x}_t - x)} \rangle_M = \int_{x \in M,\, \hat{y} \in \widehat{M}} dx\, d\hat{y}\, e^{\beta \cdot (\hat{y} - x)} \operatorname{Prob}_t(x \to \hat{y})$$

$$= \frac{1}{|M|} \int_{\substack{x \in M \\ \hat{y} \in \widehat{M}}} dx\, d\hat{y}\, e^{\beta \cdot (\hat{y} - x)} \delta(\hat{y} - \hat{\phi}^t(x)) .$$

Here, $|M| = \int_M dx$ is the volume of the elementary cell $M$. Note that there is a unique lattice translation $\hat{n}$ such that $\hat{y} = y - \hat{n}$, with $y \in M$. Therefore, and this is the main point, translational invariance together with (1.1) can be used to reduce this average to the elementary cell:

$$\langle e^{\beta \cdot (\hat{x}_t - x)} \rangle_M = \frac{1}{|M|} \int_{x,y \in M} dx\, dy\, e^{\beta \cdot (\hat{\phi}^t(x) - x)} \delta(y - \phi^t(x)) . \tag{2.1}$$

In this way the global $\hat{\phi}^t$ flow averages can be computed by following the flow $\phi^t$ restricted to the elementary cell $M$. As is well known [R], the $t \to \infty$ limit of such averages can be recovered by means of transfer operators. The Equation (2.1) suggests that we study the operator $\mathcal{L}^t$ whose kernel is given by

$$\mathcal{L}^t(y, x) = e^{\beta \cdot (\hat{x}_t - x)} \delta(y - x_t) , \tag{2.2}$$

where $\hat{x}_t = \hat{\phi}^t(x) \in \widehat{M}$, but $x, x_t, y \in M$. It is straightforward to check that this operator has the semigroup property, $\int_M dz\, \mathcal{L}^{t_2}(y, z) \mathcal{L}^{t_1}(z, x) = \mathcal{L}^{t_2 + t_1}(y, x)$. The quantity of interest (1.3) is given by the leading eigenvalue of $\mathcal{L}^t$, $\lambda_0 = e^{tQ(\beta)}$. In particular, for $\beta = 0$, the operator (2.2) is the Perron-Frobenius operator, with the leading eigenvalue $\lambda_0 = 1$ (the probability conservation).

To evaluate the spectrum of $\mathcal{L}$, consider

$$\operatorname{tr} \mathcal{L}^t = \int_M dx\, e^{\beta \cdot \hat{n}_t(x)} \delta(x - x_t) . \tag{2.3}$$

Here $\hat{n}_t(x)$ is the discrete lattice translation defined in (1.1). For discrete time and hyperbolic dynamics we have

$$\operatorname{tr} \mathcal{L}^t = \sum_{\substack{p:\sigma_p r = t, \\ r \in \mathbf{N}}} \sum_{x \in p} \frac{e^{\beta \cdot \hat{n}_t(x)}}{|\det(\mathbf{1} - \mathbf{J}^r(x))|}, \tag{2.4}$$

where the sum is over periodic points of all prime cycles $p$ whose period $\sigma_p$ divides $t$, and $\mathbf{J}_p(x) = D\phi^{\sigma_p}(x)$. Note that the sum over cycle points of $p$ can be replaced by a factor $\sigma_p$, as $\det(\mathbf{1} - \mathbf{J}_p) = \det(\mathbf{1} - \mathbf{J}_p^r(x))$ and $\hat{n}_p = \hat{n}_{\sigma_p}(x)$ are independent of $x$. For the Jacobian $\mathbf{J}_p$ this follows by the chain rule, and for the travelled distance $\hat{n}_p$ this follows by continuing the path periodically in $\widehat{M}$. For the discrete time case we finally obtain

$$\det(\mathbf{1} - z\mathcal{L}) = \prod_p \exp\left(-\sum_{r=1}^\infty \frac{z^{\sigma_p r}}{r} \frac{e^{r\beta \cdot \hat{n}_p}}{|\det(\mathbf{1} - \mathbf{J}_p^r)|}\right), \tag{2.5}$$

where the product runs over the set $\mathcal{P}$ of prime cycles.

Generalization to continuous time [Bo, CE1] amounts to the replacement $z^{\sigma_p} \to e^{-s\sigma_p}$, where $\sigma_p$ is now the (not necessarily integer) period of the prime cycle $p$:

$$Z(\beta, s) = \prod_{p \in \mathcal{P}} \exp\left(-\sum_{r=1}^\infty \frac{1}{r} \frac{e^{(\beta \cdot \hat{n}_p - s\sigma_p)r}}{|\det(\mathbf{1} - \mathbf{J}_p^r)|}\right). \tag{2.6}$$

The associated Ruelle $\zeta$ function is then (see for ex. ref. [AAC] for details)

$$1/\zeta(\beta, s) = \prod_{p \in \mathcal{P}} \left(1 - \frac{e^{\beta \cdot \hat{n}_p - s\sigma_p}}{|\Lambda_p|}\right), \tag{2.7}$$

with $\Lambda_p = \prod_e \lambda_{p,e}$ the product of the expanding eigenvalues of $\mathbf{J}_p$. (This formula can also be obtained by using the suspension formula in Appendix C.3 of [R], with the discrete dynamical system leading from collision to collision with the central disk, and the ceiling function being the time between these hits.)

Our first result is therefore: *The function $Q(\beta)$ of Eq. (1.3) is the largest solution of the equation $Z(\beta, Q(\beta)) = 0$ (or equivalently, of $1/\zeta(\beta, Q(\beta)) = 0$).*

The above infinite products can be rearranged as expansions with improved convergence properties [AAC]. To present the result, we define $t_p = e^{\beta \cdot \hat{n}_p - s\sigma_p}/|\Lambda_p|$, and expand the $\zeta$ function (2.7) as a formal power series,

$$\prod_{p \in \mathcal{P}} (1 - t_p) = 1 - \sum_{p_1, \ldots, p_k}{}' t_{\{p_1, \ldots, p_k\}}, \tag{2.8}$$

where

$$t_{\{p_1, \ldots, p_k\}} = (-1)^k t_{p_1} t_{p_2} \cdots t_{p_k},$$

and the sum is over all distinct non-repeating combinations of prime cycles. Following the derivation of Eq.(35) and Eq.(80) in [AAC] we get, for example,

$$\left.\frac{\partial}{\partial \beta_i} Q(\beta)\right|_{\beta=0} = \frac{\sum'(-1)^k (\hat{n}_{p_1} + \cdots + \hat{n}_{p_k})_i/|\Lambda_{p_1} \cdots \Lambda_{p_k}|}{\sum'(-1)^k (\tau_{p_1} + \cdots + \tau_{p_k})/|\Lambda_{p_1} \cdots \Lambda_{p_k}|}, \tag{2.9}$$

with sums as in (2.8). Two derivatives yield our second result:

*The diffusion constant (1.6) is given by*

$$D = \frac{1}{2} \frac{\sum'(-1)^k (\hat{n}_{p_1} + \cdots + \hat{n}_{p_k})^2/|\Lambda_{p_1} \cdots \Lambda_{p_k}|}{\sum'(-1)^k (\tau_{p_1} + \cdots + \tau_{p_k})/|\Lambda_{p_1} \cdots \Lambda_{p_k}|}. \tag{2.10}$$

Note that the global trajectory is in general not periodic, $\hat{n}_p \neq 0$; nevertheless, the reduction to the elementary cell enables us to compute relevant quantities in the usual way, in terms of periodic orbits.

## Consequences of lattice symmetry

The lattice symmetry of the Lorentz billiard has important consequences on the properties of the function $Q(\beta)$ which are best illustrated by introducing its analytic continuation to $\beta = ik$. The function $F(k) = Q(ik)$ is the rate associated with the incoherent scattering function $\langle \exp ik \cdot (\hat{x}_t - x) \rangle_M$ considered in light or neutron scattering experiments in liquids, in particular, by Van Hove [BY, VH]. The vector $k$ is interpreted as the wavenumber of the hydrodynamic modes of diffusion which we also find in the Lorentz gas. The function $F(k)$ turns out to be a dispersion relation since $F(k) = -Dk^2 + \mathcal{O}(k^4)$ in an isotropic diffusive system. The isotropy of a liquid implies that the dispersion relation only depends on the amplitude $|k|$ of the wavenumber.

On the other hand, the lattice symmetry of the Lorentz gas imposes special restrictions on the properties of the dispersion relation $F(k)$ and on the values taken by the wavenumber. The present classical problem of diffusion is similar to the quantum motion of a particle in a periodic potential. Hence, the wavenumber takes its values in the so-called Brillouin zone [BSW, H]. A mode of diffusion is associated with each value of the wavenumber $k$ so that the direction of $k$ is privileged in the system. As a consequence, the symmetry of the lattice is reduced by the choice of $k$. This symmetry reduction is formalized by the concept of little group associated with the wavenumber $k$, which is the subgroup of the lattice point group leaving invariant the vector $k$. For most values of $k$ inside the Brillouin zone, the little group is trivial because it only contains the identity. However, the little group is larger when the wavenumber belongs to special symmetry lines or symmetry points in the Brillouin zone. In particular, the little group coincides with the full point group when $k = 0$.

These results have the following consequences for the factorization of the zeta function $Z(k, s)$. The little group associated with $k$ has a single irreducible representation for most values of $k$ where the zeta function does not factorize. Only, at those special values of $k$ where the corresponding little group contains several irreducible representations, there exists a factorization of the zeta function of the form [CE2]

$$Z(k,s) = \prod_\alpha Z_\alpha(k,s), \qquad (3.1)$$

where $\alpha$ runs over the irreducible representations of the little group of $k$. For instance, in the triangular Lorentz gas, the Brillouin zone has several lines where the little group contains the identity together with a reflection. We can then simplify the calculation of the diffusion coefficient by following the results of Sect. 2 applied to the factor of (3.1) corresponding to the identity representation. Nevertheless, the simplification obtained by taking advantage of this reflection symmetry is modest since, in this case, the zeta function splits into only two factors.

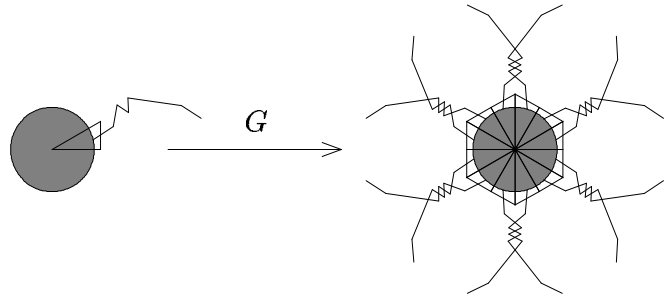

**Fig. 2**. Unfolding of a trajectory of the fundamental domain under the action of the group $G$. Note that each fundamental domain orbit corresponds to 12 distinct global orbits.

The preceding considerations concern the consequences of the lattice symmetry for the factorization of the zeta function. There is a different problem which is to express the zeta function in terms of the prime periodic orbits of the fundamental domain $\tilde{M}$ of the lattice (see Fig. 2) rather than those of the elementary (Wigner-Seitz) cell $M$. This problem is a priori independent of the factorization discussed above and presents the following difficulty. The stumbling block appears to be the non-commutativity of translations and rotations. More precisely, in contrast to (2.4), the global distance $\hat{\phi}^{r\sigma_{\tilde{p}}}(\tilde{x}) - \tilde{x}$, $\tilde{x} \in \tilde{p}$, depends on the starting cycle point if $\tilde{p}$ is only a segment of the global cycle $p$. An example is the diamond-shaped cycle of Fig. 3; depending on whether one starts at $\tilde{x}_1$ or $\tilde{x}_2$, the global distance covered in time $\sigma_{\tilde{p}}$ is either the short or the long diagonal. This difficulty can be handled in the following way.

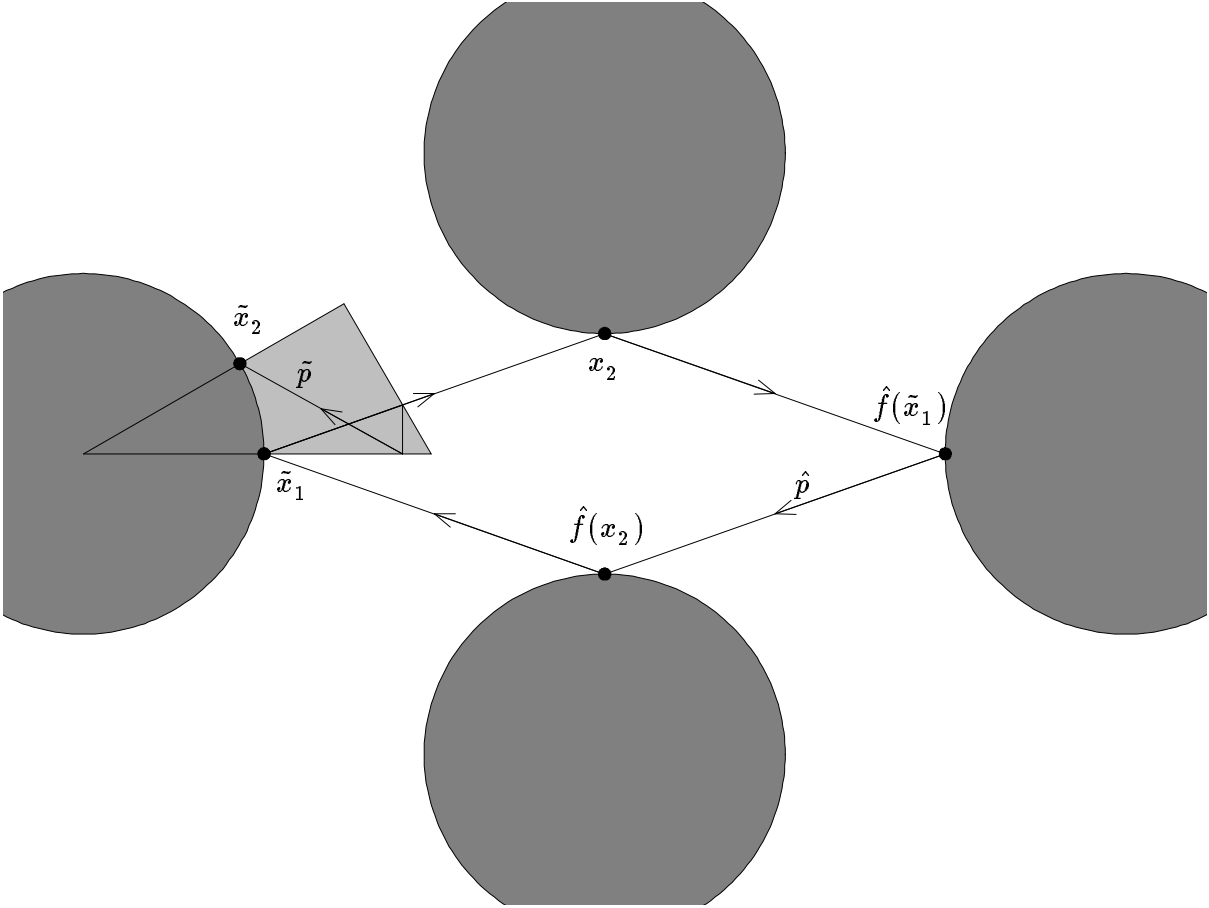

**Fig. 3**. A fundamental domain 2-cycle $\tilde{p}$ which covers in one return time only 1/2 of the corresponding global cycle $\hat{p}$. $\hat{f}(x)$ is the collision-to-collision mapping induced by the flow. To make the figure more readable, the disks have radii smaller than those needed for the "no infinite free flight" condition.

We can introduce the flow $\tilde{\phi}^t$ on the fundamental domain $\tilde{M}$ using the full point group of the lattice. Moreover, the density $\rho(x)$ on which the Perron-Frobenius operator $\mathcal{L}^t$ acts can always be decomposed using the projectors onto the irreducible representations of the full point group. For an arbitrary wavenumber $k$, the Perron-Frobenius operator will mix the different components of the density $\rho(x)$, which we can express by

$$\mathcal{L}^t(y,x) = \mathbf{R}(x;k,t)\,\delta[y - \tilde{\phi}^t(x)]\,, \tag{3.2}$$

where $\mathbf{R}(x;k,t)$ is a matrix ruling the dynamics of the different components of the density. With Eq.(3.2), the Perron-Frobenius operator is reduced to the flow in the fundamental domain $\tilde{M}$. The zeta function can then be written as a product over the prime periodic orbits $\tilde{p}$ of the fundamental domain

$$Z(k,s) = \prod_{\tilde{p} \in \tilde{\mathcal{P}}} \exp\left\{ - \sum_{r=1}^{\infty} \frac{1}{r} \, \text{tr}[\tilde{\mathbf{R}}_{\tilde{p}}(k)^r] \, \frac{\exp(-s\sigma_{\tilde{p}} r)}{|\det(1 - \tilde{\mathbf{J}}_{\tilde{p}}^r)|} \right\}, \qquad (3.3)$$

where $\tilde{\mathbf{R}}_{\tilde{p}}(k)$ is the matrix $\mathbf{R}(x;k,t)$ associated with the prime periodic orbit $\tilde{p}$. It is only when the little group of the wavenumber $k$ has nontrivial irreducible representations that the matrices $\tilde{\mathbf{R}}_{\tilde{p}}(k)^r$ split into block-diagonal submatrices which can be assigned to each irreducible representation so that the zeta function factorizes as explained with Eq.(3.2). The same discussion can be developed with $\beta = ik$.

We end this section with the remark that the signature of the lattice symmetry appears in the behavior of the function $Q(\beta_x, \beta_y)$ away from $\beta = 0$ as shown elsewhere [G] and, this, even in triangular or square lattices where diffusion is isotropic. Accordingly, the full function $Q(\beta)$ contains more information on the lattice symmetry than the diffusion matrix of the second derivatives of $Q(\beta)$.

## Conclusions

Compared to the literature [CE2, Ro], the new feature of the problem at hand is use of vector-valued functions in Eq. (1.2). The arbitrary vector $\beta$ is only a device for generating moments – the moments themselves are invariant under discrete symmetries, but it can be interpreted in terms of the wavenumber of the hydrodynamic modes of diffusion as discussed in Sect. 3.

We have thus obtained a description of global diffusive properties of an infinite periodic dynamical system, such as the Lorentz gas, in terms of periodic orbits restricted to the elementary cell. These formulas have been tested extensively in refs. [CGS, BEC] on the Lorentz gas, and in ref. [A] on 1-dimensional mappings. Related trace formulas have been independently introduced and tested numerically in ref. [V]. The formalism has been generalized to evaluation of power spectra of chaotic time series in ref. [CFP]. However, a derivation of the corresponding formulas for dynamics restricted to the fundamental domain should need further developments as sketched in Sect. 3.

In practice, the periodic orbit evaluations of the diffusion constant converge poorly compared with averages over scalar quantities such as the Lyapunov exponents. These difficulties are due to several reasons: (1) the diffusion coefficient is not a mean but a variance which is always more difficult to evaluate than mean quantities like Lyapunov exponents; (2) there is presently no simple formula for the diffusion coefficient in terms of the periodic orbits of the fundamental domain; (3) systems like the Lorentz gas do not have simple symbolic dynamics and, furthermore, the flow may have discontinuities where trajectories become tangent to the disks. These reasons affect the numerical implementation of the formulas proposed here.

*Acknowledgments*—We thank R. Artuso for communicating to us his related results on diffusion in circle maps [A] prior to publication, and to T. Schreiber for investigating in detail the applicability of the above formalism. The work reported here was performed under the auspices of the Nordita "Quantum Chaos and Measurement" workshop April–June 1991. JPE is grateful to the Niels Bohr Institute for hospitality, PC to the Carlsberg Foundation for support, and PG to the National Funds for Scientific Research (Belgium) for support.